\documentclass[12pt]{article}
\usepackage{amsmath,amssymb}
\usepackage[pdftex]{graphicx}
\usepackage{natbib}

\newcommand{\NN}{\mathbb{N}}
\newcommand{\RR}{\mathbb{R}}

\newcommand{\ZZ}{\mathbb{Z}}

\newcommand{\htau}{\tau}

\newcommand{\EE}{\mathbb{E}}
\newcommand{\PP}{\mathbb{P}}
\DeclareMathOperator{\Var}{Var}
\DeclareMathOperator{\Cov}{Cov}

\newcommand{\sd}{\,{\rm d}}
\newcommand{\rd}{{\rm d}}

\newcommand{\pred}{{\rm pred}}
\newcommand{\gen}{k}
\newcommand{\obs}{{\rm obs}}
\newcommand{\mean}{{\rm mean}}

\newcommand{\CRPS}{{\rm CRPS}}
\newcommand{\genset}{\{\text{``$\obs$''}, \text{``$\pred$''}\}}
\newcommand{\cond}{{\rm v}}

\newtheorem{theorem}{Theorem}
\newtheorem{remark}{Remark}

\begin{document}

\def\spacingset#1{\renewcommand{\baselinestretch}
{#1}\small\normalsize} \spacingset{1}

  \title{\bf Statistical Post-Processing of Forecasts for Extremes Using Bivariate Brown-Resnick Processes
       with an Application to Wind Gusts}
  \author{Marco Oesting\thanks{Marco Oesting is Postdoctoral Research Fellow, Faculty of Geo-Information Science
	                             and Earth Observation (ITC), University of Twente, Enschede, The Netherlands
															 (E-mail: m.oesting@utwente.nl). Martin Schlather is Professor, Institut f\"ur 
															 Mathematik, Universit\"at Mannheim, Germany (E-mail: schlather@math.uni-mannheim.de).
															 Petra Friederichs is Senior Lecturer, Meteorologisches Institut, Universit\"at Bonn,
															 Germany (E-mail: pfried@uni-bonn.de). 
                               This research has been supported by Volkswagen Stiftung within the project 
															 ``Mesoscale Weather Extremes -- Theory, Spatial Modeling and Prediction (WEX-MOP)". 
															 The research of M.~Oesting has also partially been funded by the ANR project 'McSim'.
															 Observational data and COSMO-DE-EPS forecasts have been kindly provided by Deutscher 
															 Wetterdienst in Offenbach, Germany.}\hspace{.2cm},
     Martin Schlather and Petra Friederichs
		}
	\date{}		
  \maketitle

\begin{abstract}
 To improve the forecasts of weather extremes, we propose a joint spatial model
 for the observations and the forecasts, based on a bivariate Brown-Resnick
 process. As the class of stationary bivariate Brown-Resnick processes is fully
 characterized by the class of pseudo cross-variograms, we contribute to the
 theorical understanding of pseudo cross-variograms refining the knowledge of
 the asymptotic behaviour of all their components and introducing a 
 parsimonious, but flexible parametric model. Both findings are of interest in 
 classical geostatistics on their own. The proposed model is applied to real 
 observation and forecast data for extreme wind gusts at 119 stations in 
 Northern Germany.
\end{abstract}

\bigskip

\section{Introduction}

Spatial extremes may occur in various forms such us heavy rainfall, floods,
heat waves or wind gusts. In view of their severe consequences, an adequate and
precise forecast of these events is of great importance. However, the rareness 
of extreme events impedes any such task and, consequently, existing forecasts 
often lack accuracy. In meteorology, for example, forecasting extreme wind 
gusts, which are defined as peak wind speeds over a few seconds, is exacerbated
by the short temporal and spatial ranges. Furthermore, numerical weather 
prediction (NWP) models provide estimates or diagnoses of wind gusts based on
empirical knowledge only \citep[cf.][]{brasseur-2001}. Although wind is a 
prognostic variable in NWP models, its values represents an average wind speed 
over a few minutes or longer depending on the grid spacing of the NWP model. 
Hence, post-processing procedures are needed that allow for an enhanced 
probabilistic forecast.
\medskip

Occurring as limits of normalized pointwise maxima of stochastic processes, 
max-stable processes provide a suitable framework for the description of 
spatial extreme events, commonly used in environmental sciences 
\citep{coles-1993, coles-tawn-1996, huser-davison-2013}. Of particular interest
is the subclass formed by Brown-Resnick processes which arise as limits of 
rescaled maxima of Gaussian processes \citep{brown-resnick-1977, kab-etal-2009,
kabluchko-2011}.

During the last years, max-stable processes have been frequently applied as 
models for spatial extremes in environmental sciences. For instance, 
\citet{engelke-etal-2012} and \citet{genton-etal-2014} recently used max-stable
processes to model extreme wind speed observations. The model we propose will 
go one step further, also taking into account the forecasts in two different 
aspects: First and in contrast to \citet{engelke-etal-2012} and 
\citet{genton-etal-2014}, we consider the mean forecast to get a normalized
version of the extreme observations. Second, besides the observable variable 
of interest itself, the corresponding forecast is included as second variable 
yielding a bivariate max-stable process. Here, we will focus on the class of
bivariate Brown-Resnick processes \citep[cf.][]{molchanov-stucki-2013, 
genton-etal-2014} to exploit the statistical relation between observable data
and the corresponding forecast. Modeling the behavior of observational data, a 
sample from the distribution of the observations conditional on the forecast is
supposed to provide more realistic results than the original forecast and thus
will appear as an appropriate probabilistic post-processed forecast.
\medskip

The paper is structured as follows: In Section \ref{sec:univ}, we present a
univariate model for extreme observations, which may, in general, provide a 
first alternative to the original forecast. We introduce a model for the 
marginal distribution, i.e.\ the distribution of the observable variable of 
interest at a single location, motivating the normalization of its extremes 
by the mean forecast. The spatial dependence structure is incorporated into 
the model by the use of univariate Brown-Resnick processes. Section 
\ref{sec:biv} is dedicated to the bivariate Brown-Resnick process which serves
as a joint model for both the maximally observed and forecasted quantities. We
deduce a necessary condition on the asymptotic behavior of the pseudo 
cross-variogram and provide a flexible cross-variogram model which leads to a 
stationary bivariate Brown-Resnick process. In Section \ref{sec:fit}, we 
describe how the model can be fitted to data. Based on this model, we propose a
post-processing procedure which is presented in Section \ref{sec:postproc}.
Further, we provide tools to verify the procedure and the underlying models.
Finally, the methods presented in Sections \ref{sec:fit} and \ref{sec:postproc}
are applied to real observation and forecast data for extreme wind gusts 
provided by the German's National Meteorological Service, Deutscher 
Wetterdienst (DWD) (Section \ref{sec:data}).
\medskip

\section{Modeling by a Univariate Random Field} \label{sec:univ}
In this section, we present a spatial model for the observed pointwise maximum
$V_{\max}^\obs$ within a specific time period. To this end, we assume that, for 
each location and time period, the maximum $V_{\max}^\obs$ is based on 
observations at $N$ equidistant instants of times per period, that is, we have
$V_{\max}^\obs = \max_{t=1,\ldots,N} V^\obs_t$ for $V_1^\obs, \ldots, V_N^\obs
\sim F_\vartheta$ for some parameter $\vartheta$. Here, the probability 
distributions $F_\vartheta$ are supposed to form a location-scale family with 
finite second moments, i.e.\  $\vartheta = (m,s) \in \RR \times (0,\infty)$ 
with $F_{(m,s)}(x) = F_{(0,1)}\left(\frac{x-m}{s}\right)$, $x \in \RR$, and 
$F_{(0,1)}$ is standardized to mean zero and unit variance.
We assume $\vartheta=(m,s)$ to
be temporally constant at each location within the same time period, but allow 
the values to vary among different locations and different time periods. The 
values of $m$ and $s$ will essentially be estimated from  the bulk of the 
distribution, not the tail, and thus, they can often be extracted accurately from 
forecasts. Within the same time period and at the same location, the observable
variables $V_1^\obs,\ldots, V_N^\obs$ are assumed to be subsequent $N$ elements
of a stationary time series $(V_t^\obs)_{t \in \ZZ}$. Furthermore, we assume that 
the standardized distribution $F_{(0,1)}$ belongs to the max-domain of 
attraction of some univariate extreme value distribution $G_\xi$, $\xi\in \RR$,
i.e.\ there are sequences $(a_n)_{n\in\NN}$, $a_n>0$, and $(b_n)_{n\in\NN}$,
$b_n \in \RR$, such that
$$ F_{(0,1)}^n(a_nx+b_n) \stackrel{n \to \infty}{\longrightarrow} G_\xi(x), \quad 1 + \xi x > 0,$$
where
\begin{equation*}
 G_\xi(x) = \begin{cases}
             \exp(-(1 + \xi x)^{-1/\xi}), & \xi \neq 0,\\
             \exp(-\exp(-x)), & \xi = 0,
            \end{cases}
\end{equation*}
for $1+\xi x>0$. As the second moment of $F_{(0,1)}$ is assumed to be finite, 
we have $\xi<0.5$. Under some conditions on the regularity and the 
dependence of the stationary sequence $V^\obs_1, V^\obs_2, \ldots$, we obtain
that
\begin{equation}
 \PP\left( \frac{ \max_{i=1, \ldots, n} V^\obs_i - m - \widetilde b_n s}{\widetilde a_n s}
   \leq x \right)  \stackrel{n \to \infty}{\longrightarrow} G_\xi(x), \quad 1 + \xi x > 0,
\end{equation}
where $\widetilde a_n = a_n \theta^{-\xi}$ and
$\widetilde b_n = b_n - \xi^{-1} (1-\theta^{-\xi})$ for some $\theta \in (0,1]$
called extremal index \citep[cf.][]{coles-2001,leadbetter-etal-1983}.

Let $m=m(l,p)$ and $s=s(l,p)$ be the mean of the variable at 
location $l$ and period $p$ and its standard deviation, respectively. Let
$$
G_{\xi,\mu,\sigma} (x) = G_\xi((x - \mu) /\sigma), \qquad 1 +
\xi(x-\mu)/\sigma > 0
$$
be the generalized extreme value distribution (GEV). Then, considering 
$V_{\max}^\obs = V_{\max}^\obs(l,p)$ for large $N$, we have approximately that
\begin{equation} \label{eq:single-loc-model-obs}
 \frac{V_{\max}^\obs(l,p) - m(l,p)}{s(l,p)} \sim G_{\xi^\obs,\mu^\obs(l),\sigma^\obs(l)}.
\end{equation}
Here, the GEV parameters are assumed to be the same for every time period,
which, in general, enables us to estimate the parameters for current and
future time periods from past data. As common in many applications, the 
extreme value index $\xi$ is also assumed to be constant in space, while
$\mu^\obs(l)$ and $\sigma^\obs(l)$ may depend on the location $l$. In contrast
to $\mu^\obs$ and $\sigma^\obs$, $m(l,p)$ and $s(l,p)$ vary in space and time 
and may be interpreted as normalizing constants that will be the same for 
observation and forcasts. As $m(l,p)$ and $s(l,p)$ are defined as mean and 
standard deviation of the variable of interest, the parameters $\mu^\obs(l)$ 
and $\sigma^\obs(l)$ are uniquely determined. Marginal transformation yields
that
\begin{equation} \label{eq:obs-gumbel}
X^\obs(l,p) = \frac 1 {\xi^\obs}
\log\left( 1 + \xi^\obs \frac{V_{\max}^\obs(l,p) - m(l,p) -
    s(l,p)  \mu^\obs(l)}
  {s(l,p)  \sigma^\obs(l)}\right)
\end{equation}
is standard Gumbel distributed for every location $l$ and time period $p$.
\medskip

Perceiving the set of locations as a subset of $\RR^2$ and the set of periods
as a subset of $\ZZ$, the transformed observations can be regarded as 
realizations of a spatio-temporal random field
$\{X^\obs(l,p),\,l\in\RR^2,\,p\in\ZZ\}$. While we assume that the spatial 
random fields $\{X^\obs(l,p),\,l\in\RR^2\}$, $p\in\ZZ$, are independent and
identically distributed, we allow for a non-trivial spatial dependence 
structure. Here, we use the class of Brown-Resnick processes 
that can be defined for arbitrary dimensions $D$
\citep{brown-resnick-1977,kab-etal-2009}: Let $\Pi=\sum_{i\in\NN} \delta_{U_i}$
be a Poisson point process on $\RR$ with intensity $e^{-u}\sd u$ and, 
independently of $\Pi$, let $W_i$, $i \in \NN$, be independent copies of a
zero-mean Gaussian random field $\{W(s),\,s\in\RR^D\}$ with stationary 
increments and semi-variogram~$\gamma(\cdot)$ defined by
$$\gamma(s) = \frac 1 2 \Var(W(s)-W(0)), \quad s\in\RR^D.$$
Then, the random field $Z$ defined by
\begin{align*}
 Z(s) = \max_{i \in \NN} \left(U_i+W_i(s)-\frac{\Var(W(s))}2\right), \quad s\in\RR^D,
\end{align*}
and called Brown-Resnick process associated to the semi-variogram~$\gamma$,
is stationary and max-stable with standard Gumbel margins and its law only
depends on the semi-variogram~$\gamma$ \citep{kab-etal-2009}.
For the application of the Brown-Resnick model to observed data with
locations in $\RR^2$, we propose to restrict to semi-variograms from a 
flexible parametric subclass, such as semi-variograms of the type
\begin{equation} \label{eq:gamma-univ}
 \gamma_{\vartheta}(h) = \| s A(b,\zeta) h\|^{\alpha}, \quad h \in \RR^2,
\end{equation}
with $\vartheta = (s, b, \zeta, \alpha)$ for $s, b > 0$,
$\zeta \in (-\pi/4,\pi/4]$ and $\alpha \in (0,2]$. Here, the matrix
$A(b,\zeta) \in \RR^{2\times2}$ allows for geometric (elliptical)
anisotropy, i.e.
\begin{equation} \label{eq:aniso}
 A(b,\zeta) = \left(   \begin{array}{cc}
                        \cos \zeta &   \sin \zeta\\
                    - b \sin \zeta & b \cos \zeta
                       \end{array}   \right)
\end{equation}
\citep[cf.][Subsection 2.5.2]{chiles-delfiner-2012}, and $s$ is an overall 
scale factor.

\section{Modeling by a Bivariate Random Field} \label{sec:biv}

In this section, we also take into account the dependence between the observed
maximum $V_{\max}^\obs$ and its forecast $V_{\max}^\pred$. As $V_{\max}^\pred$
is a forecast for $V_{\max}^\obs$, it seems reasonable to use a GEV model 
similar to the one described in Section \ref{sec:univ} with possibly different 
parameters $\xi^\pred$, $ \mu^\pred(\cdot)$ and $ \sigma^\pred(\cdot)$, i.e.
\begin{equation} \label{eq:single-loc-model-pred}
  \frac{V_{\max}^\pred(l,p) - m(l,p)}{s(l,p)} \sim G_{\xi^\pred, \mu^\pred(l), \sigma^\pred(l)}
\end{equation}
(cf.\ Equation \eqref{eq:single-loc-model-obs}). Marginally transforming 
$V_{\max}^\pred$ analogously to \eqref{eq:obs-gumbel} yields a random field
$\{X^\pred(l,p), \, l \in \RR^2, \, p \in \ZZ\}$ with standard Gumbel margins.
Thus, we end up with bivariate spatial random fields
$\{(X^\obs(l,p),X^\pred(l,p)), \, l \in \RR^2\}$ which are 
assumed to be independent and identically distributed for  $p \in \ZZ$.

A bivariate Brown-Resnick processes can be constructed in the following way
\citep{molchanov-stucki-2013, genton-etal-2014}: 
Let $\sum_{i \in \NN} \delta_{U_i}$ be a Poisson point process
on $\RR$ with intensity measure $e^{-u} \sd u$. Further, let $W_i$, $i\in\NN$,
be independent copies of a bivariate zero mean Gaussian process
$W = (W^{(1)}, W^{(2)})^\top = \{(W^{(1)}(s), W^{(2)}(s))^\top:\,s \in \RR^D\}$
such that the so-called pseudo cross-variogram \citep{CBH89,PKW93},
$\gamma(h) = (\gamma_{ij}(h))_{i,j \in \{1,2\}}$ defined by
\begin{equation*}
 \gamma_{ij}(h) = \frac12 \Var(W^{(i)}(s+h) - W^{(j)}(s)), \quad h \in \RR^D,
\end{equation*}
does not depend on $s \in \RR^D$. Analogously to the univariate Brown-Resnick 
process, it can be shown that the bivariate Brown-Resnick process
$Z = (Z^{(1)},Z^{(2)})^\top$ defined by
\begin{equation} \label{eq:def-bivBR}
Z^{(j)}(s) = \max_{i \in \NN} \left(U_i + W_i^{(j)}(s) - \frac{\Var(W^{(j)}(s))} 2\right),
\quad s \in \RR^D, \quad j=1,2,
\end{equation}
is max-stable and stationary and its law only depends on the pseudo 
cross-variogram $\gamma$.
\medskip

\begin{remark}
 The fact that $(\gamma_{ij}(h))_{i,j=1,2}$ can be defined independently of
 $s \in \RR^D$ implies that $W$ is intrinsically stationary, i.e.\ the process
 $\{W(s+h)-W(s):\,s \in \RR^D\}$ is stationary for every $h \in \RR^D$. Both
 conditions, however, are not equivalent.
\end{remark}

Indeed, \cite{molchanov-stucki-2013} already gave necessary and sufficient
conditions for a multivariate process of Brown-Resnick type to be stationary.
For a fixed intensity $e^{-u} \sd u$ of the Poisson point process, the 
conditions on Gaussian processes given in Theorem 5.3 in 
\cite{molchanov-stucki-2013} can be shown to be equivalent to the conditions on
the process $W$ stated above (if we additionally require $Z$ to have standard
Gumbel margins) by a straightforward computation. Thus, the Gaussian processes 
in the above definition of bivariate Brown-Resnick processes are essentially the
only ones that yield a stationary max-stable process.
\medskip

In the following, we investigate the structure and the asymptotic behavior of
bivariate variograms that are translation invariant, refining the result by 
\citet{PKW93} that $\lim_{h \to \infty} \gamma_{12}(h) / \gamma_{11}(h) = 1$
if $\gamma_{11}$ is unbounded. This allows us to find valid models for 
bivariate Brown-Resnick processes. The following theorem, as well as the
statements above, immediately extend to the general multivariate case.
The proof is given in the Appendix.

\begin{theorem} \label{thm:restriction}
 Let $W = (W^{(1)}, W^{(2)})^\top$ be a bivariate second-order process on
 $\RR^D$ with pseudo cross-variogram $(\gamma_{ij}(h))_{i,j \in \{1,2\}}$,
 defined by $\gamma_{ij}(h) = \frac 1 2 \Var(W^{(i)}(s+h) - W^{(j)}(s))$
 which does not depend on $s \in \RR^D$.
 Then, we have
 \begin{equation*}
   \sqrt{\gamma(h)} = \sqrt{(\gamma_{ij}(h))_{i,j \in \{1,2\}}}
                    = \Bigg( \begin{array}{cc}
                          1 & 1\\ 1 & 1 \end{array}\Bigg) \sqrt{\gamma_0(h)}
                 +  \Bigg( \begin{array}{cc}
                          f_{11}(h) & f_{12}(h)\\
                          f_{21}(h) & f_{22}(h)
                          \end{array}\Bigg)
 \end{equation*}
 for some univariate variogram $\gamma_0$ and bounded functions
 $f_{11}, f_{12}, f_{21}, f_{22}: \RR^D \to \RR$.
\end{theorem}

As the components of a translation invariant bivariate pseudo cross-vario\-gram
only differ by a function that may increase only with a rate of order 
$O(\sqrt{\gamma_0(h)})$ (Theorem \ref{thm:restriction}), a reasonable and not 
too restrictive model for the corresponding bivariate Gaussian random field 
$W =(W^{(1)},W^{(2)})^\top$ is given by
\begin{equation*}
  W(s) = \Bigg( \begin{array}{c} 1\\ 1 \end{array}\Bigg) V_1(s) + V_2(s),
  \quad s \in \RR^D,
\end{equation*}
where $V_1$ is a univariate Gaussian random field with stationary increments and
semi-variogram $\gamma_0$ and $V_2$ is a bivariate stationary Gaussian random
field with cross-covariance function $C(h) = (C_{ij}(h))_{i,j \in \{1,2\}}$,
independent from $V_1$. Then, the pseudo cross-variogram $\gamma$ of $W$ has the form
\begin{align*}
 \gamma_{ij}(h) ={}& \gamma_0(h) + \frac 1 2 C_{ii}(0) + \frac 1 2 C_{jj}(0) - C_{ij}(h),
 \quad i,j \in \{1,2\}, \ h \in \RR^D.
\end{align*}
Analogously to the univariate case, we propose to restrict to a parametric
subclass of semi-variograms for $\gamma_0$ such as
\begin{align*}
 \gamma_0(h) = \sigma^2 \frac{(\kappa \|h\|)^2}{((\kappa \|h\|)^2+1)^\beta}
\end{align*}
where $\sigma, \kappa > 0$ and $\beta \in (0,1)$. Here, $\gamma_0$ is a valid
univariate variogram as $h \to \|h\|^2$ is a variogram and 
$\lambda \mapsto \lambda/(\lambda+1)^\beta$ is a Bernstein function
\citep[cf.][]{berg-etal-1984,schilling-etal-2010}. Note that $\gamma_0$
is a variogram of power law type modified to be smooth at the origin. For the 
bivariate cross-covariance $C$, we propose to use the parsimonious bivariate 
Mat\'ern model \citep[cf.][]{gneiting-etal-2010}, which is a bivariate 
generalization of one of the most widely used models in geostatistics, the 
Mat\'ern model \citep[cf.][for example]{guttorp-gneiting-2006,stein-1999}. 
To increase the flexibility of the model, we further add a spatially constant 
effect with variance $c^2$ in the second component. Thus, $C$ has the form
\begin{align*}
 C_{11}(h) ={}& \phantom{c^2+} \sigma_1^2 \frac{2^{1-\nu_1}}{\Gamma(\nu_1)} (a \|h\|)^{\nu_1} K_{\nu_1}(a\|h\|),  \displaybreak[0]\\
 C_{12}(h) {}={} C_{21}(h) 
    ={}& \rho \sigma_1 \sigma_2 \frac{2^{1-\nu_{12}}}{\Gamma(\nu_{12})} (a \|h\|)^{\nu_{12}} K_{\nu_{12}}(a\|h\|),\displaybreak[0]\\
 C_{22}(h) ={}& c^2 + \sigma_2^2 \frac{2^{1-\nu_2}}{\Gamma(\nu_2)} (a \|h\|)^{\nu_2} K_{\nu_2}(a\|h\|),
\end{align*}
where $\sigma_1,\sigma_2,c \geq 0$, $a>0$, $\nu_1,\nu_2>0$, $\nu_{12} = 
(\nu_1+\nu_2)/2$ and $|\rho| \leq 2 (\nu_1 \nu_2)^{1/2}/(\nu_1+\nu_2)$. Note 
that as the common summand $\gamma_0$ is smooth at the origin, the behavior of
$\gamma_{ii}$ near the origin, i.e.\ the differentiability of $W^{(i)}$, 
depends only on the behavior of $C$ which can be modeled flexibly by the smoothness
parameters $\nu_1$ and $\nu_2$ of the bivariate Mat\'ern model. In particular 
we have, as $\|h\| \to 0$ and for some $k(\nu) >0$, that
\begin{equation*}
\gamma_{ii}(h) = \begin{cases}
                   k(\nu_i) (a\|h\|)^{2\nu_i} + O(\|h\|^2), & \nu_i < 1,\\
                   k(1) (a\|h\|)^2 \log\|h\| + O(\|h\|^2), & \nu_i = 1,\\
                   k(\nu_i) (a\|h\|)^2 + o(\|h\|^2), & \nu_i >1
                 \end{cases}
\end{equation*}
\citep[cf.][]{stein-1999}. Furthermore, the sample paths are $m$ times 
differentiable if and only if $\nu > m$ \citep{gelfand-etal-10}. The behavior
of the $\gamma_{ii}$ as $\|h\| \to \infty$, which has to be the same for all 
components by Theorem \ref{thm:restriction}, is parameterized by $\beta$ as we
have $\gamma_{ii}(h) \|h\|^{-2(1-\beta)}\rightarrow 1$ as 
$\|h\|\rightarrow \infty$.
To increase the applicability of our model to real data located in $\RR^2$, we
further allow for geometric anisotropy, replacing $\|h\|$ by $\|A(b,\zeta) h\|$
where $A(b,\zeta)$ is the anisotropy matrix defined in \eqref{eq:aniso}. Thus, 
we obtain the variogram model $\gamma(\vartheta;\cdot)$ given by
\begin{align}
 \gamma_{ii}(\vartheta;h) ={}& \sigma^2 \frac{(\kappa\|A(b,\zeta)h\|)^2}{((\kappa\|A(b,\zeta)h\|)^2+1)^\beta}
                              + \sigma_i^2 \left(1 - \frac{2^{1-\nu_i}}{\Gamma(\nu_i)}
                                 (a \|A(b,\zeta)h\|)^{\nu_i} K_{\nu_i}(a\|A(b,\zeta) h\|)\right), \nonumber \displaybreak[0]\\
 \gamma_{12}(\vartheta;h) ={}& \sigma^2 \frac{(\kappa\|A(b,\zeta)h\|)^2}{((\kappa\|A(b,\zeta)h\|)^2+1)^\beta} + 
                               \frac{\sigma_1^2 + c^2 + \sigma_2^2} 2 \nonumber\\
                             & - \rho \sigma_1 \sigma_2
                       \frac{2^{1-\nu_{12}}}{\Gamma(\nu_{12})} (a \|A(b,\zeta) h\|)^{\nu_{12}} K_{\nu_{12}}(a\|A(b,\zeta) h\|), \label{eq:bivario}
\end{align}
for $i=1,2$ and $h \in \RR^2$ where
$\vartheta = (\sigma,\kappa,b,\zeta,\beta,c,\sigma_1,\nu_1,\sigma_2,\nu_2,a,\rho)$. 

\section{Model Fitting} \label{sec:fit}

In the following, we will assume that data $v_{\max}^\obs(l_i,p)$ and
$v_{\max}^\pred(l_i,p)$ for the maximal observed and forecasted variable of 
interest at stations $l_i$, $i=1, \ldots, n_l$ and time period $p=1,\ldots,n_p$
are available.

\subsection{Fitting of the Univariate Model}

Let henceforth be $\gen\in \genset$. We concentrate here on the estimation of
the GEV and max-stable parameters assuming that the unknown mean $m(l_i,d)$ and
standard deviation $s(l_i,p)$ of the underlying distribution $F$ have already 
been estimated by $\hat m(l_i,p)$ and $\hat s(l_i,p)$, respectively. An example
for the later estimates can be found in Section \ref{sec:data}.

Given the estimates $\hat m(l_i,p)$ and $\hat s(l_i,p)$, we obtain the 
standardized data
\begin{equation} \label{eq:std}
  y^\gen(l_i,p) = \frac{v^\gen_{\max}(l_i,p) - \hat m(l_i,p)}{\hat s(l_i,p)},
 \qquad i=1, \ldots, n_l, \ p=1, \ldots, n_p,
\end{equation}
which are assumed to be GEV distributed with parameters $\xi^\gen$,
$\mu^\gen(l_i)$ and $ \sigma^\gen(l_i)$. The parameters can be estimated via 
maximum likelihood separately for each station. As the the standardized data 
$y^\gen$ are assumed to be temporally independent, by \cite{smith-1985}, the
maximum likelihood estimators $(\hat \xi^\gen(l_i)$, $1 \leq i \leq n_l$,
are asymptotically normally distributed if $\xi^\gen > -0.5$. Thus, under the
hypothesis that
$\hat{\xi}^\gen = \frac 1 {n_l} \sum_{i=1}^{n_l} \hat\xi^\gen(l_i)$ is the true
shape parameter of the GEV at each station, the standardized residuals
$$\frac{\hat \xi^\gen(l_1) - \hat \xi^\gen}{(\widehat \Var(\hat \xi^\gen(l_1)))^{1/2}}, \ldots,
  \frac{\hat \xi^\gen(l_{n_l}) - \hat \xi^\gen}{(\widehat \Var(\hat \xi^\gen(l_{n_l})))^{1/2}}$$
are approximately standard normally distributed, where
$\widehat \Var(\hat \xi^\gen(l_i))$ is the variance of $\hat \xi^\gen(l_i)$
estimated via the Hesse matrix of the log-likelihood function. Thus, the three
hypotheses that the shape parameter, the location and the scale parameter are
spatially constant can be checked indirectly via one-sample Kolmogorov-Smirnov
tests of the corresponding residuals for the standard normal distribution.
Here, although the data for different locations may be dependent, we assume 
that the normalized estimated parameters are independent.
\medskip

By transformation \eqref{eq:obs-gumbel}, the estimates
$\hat \xi^\gen$, $\hat \mu^\gen(l_i)$ and $\hat \sigma^\gen(l_i)$
yield normalized data
\begin{equation} \label{eq:data-obs-std}
x^\gen(l_i,p) = \frac 1 {\hat\xi^\gen} \log\left(1 + \hat\xi^\gen \frac{y^\gen(l_i,p) - \hat\mu^\gen(l_i)}{\hat\sigma^\gen(l_i)}\right), \quad 1 \leq i \leq n_l, \ 1 \leq p \leq n_p.
\end{equation}
As a goodness-of-fit test of the marginal model, these can be compared to a 
standard Gumbel distribution via Kolmogorov-Smirnov tests separately for each 
station.
\medskip

In order to capture the spatial dependence structure, a univariate 
Brown-Resnick process associated to a variogram $\gamma^\gen$ as defined in
\eqref{eq:gamma-univ} is fitted to the transformed data 
$(x^\gen(l_i,p))_{1 \leq i \leq n_l, 1 \leq p \leq n_p}$. Note that there exist
numerous methods of inference for Brown-Resnick processes, see, for example,
\citet{engelke-etal-2012} for a comparison of different estimators. The method
we will use is based on the extremal coefficient function 
\citep{schlather-tawn-2003}. For a stationary Brown-Resnick process associated 
to the semi-variogram $\gamma^\gen$, the extremal coefficient function is given 
by
\begin{equation} \label{eq:theta-BR}
 \theta^\gen(s_1,s_2) = \frac{\log \PP(X^\gen(s_1) \leq x, X^\gen(s_2) \leq x)}{\log \PP(X^\gen(s_1) \leq x)}
             = 2\Phi\left(\sqrt{\gamma^\gen(s_1-s_2)/2}\right), \quad s_1,s_2 \in \RR^2,
\end{equation}
where $\Phi$ denotes the standard normal distribution function
\citep[cf.][]{kab-etal-2009}. This relation can be used for fitting 
Brown-Resnick processes to real data as the extremal coefficients 
$\theta^\gen(s_1,s_2)$ can be estimated well via the relation
\begin{equation} \label{eq:fmado-theta}
 \theta^\gen(s_1,s_2) = \frac{1 + 2 \nu^{F,\gen}(s_1,s_2)}{1 - 2\nu^{F,\gen}(s_1,s_2)}, \quad s_1,s_2 \in \RR^2,
\end{equation}
where the $F$-madogram $\nu^{F,\gen}(s_1,s_2)$ is defined by
\begin{equation} \label{eq:def-fmado}
 \nu^{F,\gen}(s_1,s_2) = \frac 1 2 \EE\left| F(X^\gen(s_1)) - F(X^\gen(s_2)) \right|, \quad s_1,s_2 \in \RR^2,
\end{equation}
and $F$ is the marginal distribution function of $X^\gen(s)$ 
\citep{cooley-etal-2006}. Thus, we obtain a plug-in estimator 
$\hat\theta^\gen(l_i,l_j)$ for the extremal coefficients 
$\theta^\gen(l_i,l_j)$, by replacing $\nu^{F,\gen}$ in \eqref{eq:fmado-theta} 
by an estimator  $\hat \nu^{F,\gen}(l_i,l_j)$, $1 \leq i,j \leq n_l$. In order
to avoid propagation of errors in marginal modeling, we choose the 
non-parametric estimator
\begin{equation} \label{eq:est-fmado}
 \hat \nu^{F,k}(l_i,l_j) = \frac 1 {2 \cdot n_p \cdot (n_p - 1)} \sum_{p=1}^{n_p} 
                           \left|R_p(x^\gen(l_i,\cdot)) - R_p(x^\gen(l_j,\cdot))\right|
\end{equation}
where $R_p(x)$ denotes the rank of the $p$-th component of some vector $x$
\citep[cf.][]{ribatet-2013}. Then, the corresponding variogram parameter
vector $\hat \vartheta^\gen$ can be estimated by a weighted least squares
fit of $\hat \theta^\gen(l_i,l_j)$ to $\theta^\gen(l_i,l_j)$ as given in
\eqref{eq:theta-BR}. As proposed by \citet{smith-1990}, we choose weights
that depend on the (estimated) variance $\widehat \Var(\theta^\gen(l_i,l_j))$
of the estimator $\theta^\gen(l_i,l_j)$. Thus, we obtain the estimator
\begin{equation}\label{eq:lsfit-gamma}
 \hat \vartheta^\gen = \arg\min_{\vartheta} \sum_{1 \leq i < j \leq n_l}
   \left( \frac{\hat \theta^\gen(l_i,l_j) - 2\Phi\left(\sqrt{\gamma^{\gen}(l_i-l_j) / 2}\right)}{\sqrt{\widehat \Var(\theta^\gen(l_i,l_j))}}\right)^2.
\end{equation}
We will further discuss the estimation of the variance of $\theta^\gen(l_i,l_j)$
in Section \ref{sec:data}.

\subsection{Fitting of the Bivariate Model}

For fitting the bivariate Brown-Resnick process
$\{(X^\obs(l), X^\pred(l))^\top: \ l \in \RR^2\}$
we consider the extremal coefficients $\theta^{k_1,k_2}(s,t)$ of $X^{k_1}(s)$
and $X^{k_2}(t)$ for $k_1,k_2 \in \genset$. They can be estimated from the
transformed data $x^\obs(l_i,p)$ and $x^\pred(l_i,p)$, $1 \leq i \leq n_l$,
$1 \leq p \leq n_p$, in the same way as in the univariate case. The resulting
estimates $\hat \theta^{k_1, k_2}(l_i,l_j)$, $1 \leq i,j \leq n_l$,
$k_1,k_2 \in \genset$ are compared to the corresponding extremal coefficients
of a bivariate Brown-Resnick process associated to the variogram 
$\gamma(\vartheta; \cdot)$ yielding the weighted least squares fit
\begin{equation*}
\hat \vartheta = \arg \min_{\vartheta} \sum_{1 \leq i,j \leq n_l}
     \sum_{k_1, k_2 \in \genset}
     \left( \frac{\hat \theta^{k_1,k_2}(l_i,l_j) - 2 \Phi(\sqrt{\gamma_{k_1,k_2}(\vartheta;l_i-l_j)/2})}
                 {\widehat \Var(\theta^{k_1,k_2}(l_i,l_j))}
     \right)^2.
\end{equation*}

\section{The Post-Processing Procedure} \label{sec:postproc}

As the bivariate Brown-Resnick process model developed in this paper describes
the joint distribution of the observed and forecasted maxima of the variable of
interest, it allows for some spatial post-processing of the original forecast.
In this section, we will describe the resulting post-processing procedure in 
more detail and provide some tools to verify the procedure and the underlying 
model.

\subsection{Post-Processing via Conditional Simulation} \label{subsec:post}

Let $\hat\xi^\obs$, $\hat\mu^\obs(\cdot)$, $\hat\sigma^\obs(\cdot)$, 
$\hat\xi^\pred$, $\hat\mu^\pred(\cdot)$, $\hat\sigma^\pred(\cdot)$ and 
$\hat \vartheta$ be estimates for the GEV and variogram parameters derived from
past training data. Further, assume that we have $v_{\max}^\pred(l_i,p)$, 
$\hat m(l_i,p)$ and $\hat s(l_i,p)$, $i=1,\ldots,n_l$, based on forecasts for 
$n_l$ locations $l_1,\ldots,l_{n_l}$ and a time period $p$ in near future. 
Then, we obtain an arbitrary number $K$ of realizations 
$(v_j(l_i))_{1 \leq i \leq n_l}$, $j=1,\ldots,K$, of the modeled distribution of 
the maximal observation conditional on the forecast by the following three-step 
procedure:

\begin{enumerate}
 \item[1.] Transform $v_{\max}^\pred(\cdot,d)$ to standard Gumbel margins:
  $$x^\pred(\cdot) = \frac 1 {\hat \xi^\pred} \log\left(
        1 + \hat \xi^\pred \frac{v_{\max}^\pred(\cdot,p) - \hat \mu_{\cond }^\pred(\cdot,p)}
                               {\hat \sigma_{\cond }^\pred(\cdot,p)} \right),$$
  where $\hat \mu^\pred_{\cond}$ and $\hat \sigma^\pred_{\cond}$ are given by
  Equation \eqref{eq:gev-gen-cond} for $k=\pred$.
 \item[2.] Conditional simulation of a bivariate Brown-Resnick process given 
  its second component: Simulate $K$ independent realizations 	
	$(x^\obs_{j}(\cdot), x^\pred_j(\cdot))$,  $j=1,\ldots,K$, of a bivariate 
	Brown-Resnick process associated to the pseudo cross-variogram
  $\gamma(\hat \vartheta^\obs; \cdot)$ with standard Gumbel margins
  conditional on $x^\pred_j(\cdot) = x^\pred(\cdot)$.
 \item[3.] Transform $x^\obs_{j}(\cdot)$ to GEV margins:
  For $j=1, \ldots, K$, set
  $$ v_{j}(\cdot,p) = \hat \sigma^\obs_{\cond }(\cdot,p)
     \frac{\exp(\hat \xi^\obs x^\obs_j(\cdot)) -1}{\hat \xi^\obs} + \hat \mu^\obs_{\cond }(\cdot,p),$$
  where $\hat\mu^\obs_{\cond}$ and $\hat\sigma^\obs_{\cond}$ are given by
  Equation \eqref{eq:gev-gen-cond} for $k=\pred$.
\end{enumerate}

The random fields obtained by this three-step procedure can be interpreted as 
post-processed probabilistic forecasts for the maxima of the variable of 
interest. While the first and the third steps only consist of marginal 
transformations, the conditional simulation in the second step is the 
challenging part of the procedure. For this step, the algorithm by 
\citet{dombry-etal-2013} can be used. Note that the algorithm, which has 
originally been designed for conditional simulation of univariate Brown-Resnick
processes, can directly be transferred to the multivariate case by perceiving 
the multivariate processes as univariate processes on a larger index set. 
However, the computations will be computationally expensive, in particular if 
the number of conditioning locations gets large.

\subsection{Verification} \label{subsec:verif}

In practical applications, the proposed post-processing procedure and the
underlying model need to be verified. Here, we do not only intend the full
bivariate Brown-Resnick model which forms the base of the post-processing
procedure, but also intermediate models such as the marginal GEV model and the 
univariate model. This allows us to evaluate the effect of incorporating the 
spatial dependence structure and the forecasted maxima, respectively.

For the evaluation and verification the different models, we choose the 
(negatively oriented) energy score \citep[cf.][]{gneiting-raftery-2007}:
$$ ES(F,x) = \int_{\RR^m} \|y-x\|^\chi \, F(\rd y)
           - \frac 1 2 \int_{\RR^m} \int_{\RR^m} \|y_1-y_2\|^\chi \, F(\rd y_1) \, F(\rd y_2),$$
where $F$ is a $\RR^m$-valued distribution, $x \in \RR^m$ is an observation,
$\chi \in (0,2)$ and $\|\cdot\|$ denotes the Euclidean norm on $\RR^m$. The 
energy score is a strictly proper scoring rule, i.e.\
$\int ES(F,x) F({\rm dx}) \leq \int ES(G,x) F({\rm dx})$
for all distribution functions $F$ and $G$ with finite moments of order $\xi$
and equality if and only if $F=G$. This indicates that the mean energy score 
for different observations is the smaller, the better the predicted 
distribution $F$ fits to the true distribution of the observation data.
Here, we will restrict ourselves to the case $\chi=1$. If $F$ is additionally a
univariate distribution, i.e.\ $m=1$, the energy score is also called 
continuous ranked probability score (CRPS). 
\medskip

By fitting the GEV parameters according to Section \ref{sec:fit}, we obtain the
following  marginal model for the maximum at location $l_i$ within time period 
$p$:
\begin{align} \label{eq:gen-model}
 & V^\gen_{\max}(l_i,p) \sim G_{\hat \xi^\gen, \hat \mu^\gen_{\cond }(l_i,p), \hat \sigma^\gen_{\cond }(l_i,p)}, \displaybreak[0]\\
\text{where} \quad &
 \hat \mu^\gen_{\cond }(l,p) {}={} \hat m(l,p) + \hat s(l,p) \hat \mu^\gen(l)
 \quad \text{and} \quad \hat \sigma^\gen_{\cond }(l,p) {}={} \hat s(l,p) \hat \sigma^\gen(l). \label{eq:gev-gen-cond}
\end{align}
First, we evaluate the improvement in predictive quality by fitting the GEV to 
the observations instead of the forecast, and thus compare $\CRPS^\obs(l_i)$ 
and $\CRPS^\pred(l_i)$ where
\begin{align*}
  \CRPS^\gen(l_i) ={}& \frac 1 {n_d} \sum_{d=1}^{n_d} \CRPS(G_{\hat\xi^\gen,\hat \mu^\gen_{\cond}(l_i,d),
    \hat \sigma^\gen_{\cond}(l_i,d)}, v_{\max}^\obs(l_i,d)).
\end{align*}
for every station $l_i$, $1 \leq i \leq n_l$, and $\gen\in\genset$. For the 
calculation, we employ the closed formula for the CRPS of a GEV provided by 
\citet{friederichs-thorarinsdottir-2012}. For $\xi \neq 0$, they obtain
\begin{equation} \label{eq:crps-gev}
 \CRPS(G_{\xi,\mu,\sigma},x) = \left(x - \mu + \frac \sigma \xi\right) (2F(x)-1)
           - \frac \sigma \xi \Big(2^\xi \Gamma(1-\xi) - 2 \Gamma_l(1-\xi,-\log F(x))\Big)
\end{equation}
where $\Gamma_l$ is the lower incomplete gamma function. 

Furthermore, the CRPS for the GEV fitted to the observations can be compared
with the CRPS of the original forecast 
\begin{align*}
 \CRPS^{{\rm orig}}(l_i) ={} \frac 1 {n_p} \sum_{p=1}^{n_p} \CRPS(F^{{\rm orig}}_{l_i,p}, v_{\max}^\obs(l_i,p))
\end{align*}
where $F^{\rm orig}_{l_i,p}$ denotes the distribution of the original 
(probabilistic) forecast for the maximum of the variable of interest at
location $l_i$ within time period $p$. If this forecast is given by an ensemble
of values, such as the output of a numerical weather prediction model, for
example, $F^{{\rm orig}}_{l_i,p}$ corresponds to the empirical distribution
function of this sample. If the forecast corresponds to a single value, 
$\CRPS^{{\rm orig}}(l_i)$ reduces to the mean absolute error.
\medskip

For verification of the univariate Brown-Resnick model as a model for the 
spatial dependence structure, we propose to compare energy scores for the
Brown-Resnick process with those of independent
$G_{\hat\xi^\obs,\hat\mu^\obs_{\cond}(l_i,d),\hat\sigma^\obs_{\cond}(l_i,d)}$
random variables. As we often do not have closed forms for the energy scores of
the higher-dimensional marginal distributions, these cannot be calculated 
exactly but need to be approximated replacing the multivariate distribution $F$
by an empirical distribution generated by simulations. We will denote the 
estimated energy scores belonging to the joint distribution at locations 
$l_{i_1},\ldots,l_{i_n}$ by $\widehat {ES}^{\rm BR}(l_{i_1},\ldots,l_{i_n})$
for the Brown-Resnick process, and $\widehat {ES}^{\rm ind}(l_{i_1},\ldots,
l_{i_n})$ for the independence model, respectively.

Finally, the full bivariate model and, thus, the quality of the proposed
post-processing procedure can be evaluated and verified by considering the CRPS
\begin{equation*}
 \CRPS^{\rm biv}(l_i) = \frac 1 {n_p} \sum_{p=1}^{n_p} \CRPS\left(F_{l_i,p \mid v_{\max}^{\pred}}, v_{\max}^\obs(l_i,p)\right)
\end{equation*}
where $F_{l_i,p \mid v_{\max}^{\pred}}$ denotes the distribution of the
observed maximum at location $l_i$, $1 \leq i \leq n_l$ within time period $p$
conditional on $v_{\max}^{\pred}$, i.e.\ the distribution of the post-processed 
forecast, with the CRPS of the original forecast, $\CRPS^{\rm orig}(l_i)$.

\section{Application to Real Data} \label{sec:data}

In this section, we will apply the fitting and verification procedure described
in Section \ref{sec:fit} to real wind gust data consisting both of observation 
and forecast data. We will see that, even though the marginal distributions are
fitted quite well, a forecast based on the single GEV for the observations is 
not able to outperform the forecast by the numerical weather prediction model. 
However, the results for the bivariate model indicate that the post-processing 
procedure proposed in Subsection \ref{subsec:post} may improve the predictive 
quality.

\subsection{The Data}

We consider observed as well as forecasted wind speed data provided by 
Germany's National Meteorological Service, the Deutscher Wetterdienst (DWD). We
use observations from 218 DWD weather stations over Germany at 360 days from 
March 2011 to February 2012. The weather stations register mean and maximum 
wind speed on an hourly basis. Due to the inertia of the measuring instruments,
the maximum wind speed approximately corresponds to the highest $3$-second 
average wind speed. Here, we use the maximum wind speed $v^\obs_{\max}(l,d)$ 
between 08 UTC and 18 UTC for each station $l$ and each day $d$.

Furthermore, for each day, forecasts for the wind speed maxima and for the
hourly mean wind speed both in 10m height above ground and for the 
$10$-hour-period from 08 UTC to 18 UTC are available. The forecasts are 
provided by the COSMO-DE ensemble prediction system (EPS) operated by DWD.
COSMO-DE \citep{baldauf-etal-2011} is a non-hydrostatic limited-area numerical
weather prediction model that gives forecasts for the next 21 hours on a 
horizontal grid with a width of 2.8km covering Germany and neighboring 
countries. For each variable of interest, the COSMO-DE EPS yields forecasts
consisting of 20 ensemble members stemming from COSMO-DE runs with five 
different physical parameterizations and four different lateral boundary 
conditions provided by global model forecasts. For more details on the
Consortium for Small-scale Modeling see {\tt http://www.cosmo-model.org/}, and 
\cite{gebhardt-etal-2011} and \cite{peralta-etal-2012}, for COSMO-DE EPS.

The COSMO-DE EPS is initialized every 3 hours. Here, we take the forecasts that
are initialized at 00 UTC. If we use the forecasts for the nearest grid 
location of a station, we obtain the forecasts
$v_{\mean}^{(1)}(l,d,\htau)$, $\ldots$, $v_{\mean}^{(20)}(l,d,\htau)$,
$\htau \in \{9, 10, \ldots, 18\}$, and $v_{\max}^{(1)}(l,d), \ldots, v_{\max}^{(20)}(l,d)$
for every weather station $l$ and every day $d$. Here, $v_{\mean}^{(j)}(l,d,\htau)$
and $v_{\max}^{(j)}(l,d)$ denote the forecast for the mean wind speed between
$(\htau-1)$ UTC and $\htau$ UTC and the maximal wind speed, respectively, at station $l$
and day $d$, forecasted by the $j$th COSMO-DE ensemble member.
\medskip

For the application of our model with a stationary spatial dependence structure,
in the following, we will restrict ourselves to forecasted and observed data
for 119 DWD stations north of $51^\circ$N as the northern part of Germany has a
much more homogeneous topography than the southern part. We will denote the
locations of these stations by $l_1, \ldots, l_{119}$.

\subsection{Applying the Univariate Model} \label{subsec:apply-univ}

As the wind speed observations correspond to $3$-second averages, the daily 
maximal wind gusts $v_{\max}^{\obs}$ can be perceived as the maximum of a 
long time series. Further, the distribution of a single wind speed is 
frequently modeled by a Weibull or a Gamma distribution
\citep[e.g.,][]{conradson-etal-1984,pavia-etal-1986,sloughter-etal-2007},
that is, the single observations may be assumed to come from a location-scale
family of distributions provided that the shape parameter is spatially and
temporally constant. These considerations give support to the usage of the GEV
model presented in Section \ref{sec:univ} as a model for the maximal wind speed
$V^\gen_{\max}(l_i,d)$, at station $l_i$, $i \in \{1,\ldots,119\}$, and day 
$d \in \{1,\ldots,360\}$. For fitting a GEV distribution to the standardized
wind speeds $y^\gen(l_i,d)$ as defined in \eqref{eq:std}, we need the estimates
$\hat m(l_i,d)$ and $\hat s(l_i,d)$ corresponding to the mean and the standard 
deviation of the underlying wind speed distribution. Here, instead of direct 
estimates for these characteristics, we use 
\begin{align} 
             \hat m(l_i,d) ={}& \max_{j=1}^{20} \frac 1 {10} \sum\nolimits_{\htau=9}^{18} v_{\mean}^{(j)}(l_i,d,\htau) 
						\label{eq:v-mean} \displaybreak[0]\\
 \text{and } \hat s(l_i,d) ={}& \left(\frac 1 {199} \sum\nolimits_{j=1}^{20} \sum\nolimits_{\htau=9}^{18} (v_{\mean}^{(j)}(l_i,d,\htau) - \hat m(l,d))^2\right)^{1/2} \label{eq:v-sdev}.
\end{align}
Even though not providing consistent estimates for mean and standard deviation,
$\hat m(l_i,d)$ and $\hat s(l_i,d)$ ensure that $y^\gen(l_i,d)$ is invariant
under affine transformations of the underlying distribution as long as the
transformation is reflected in the forecasts $v_{\mean}^{(j)}$. This choice
of $\hat m(l_i,d)$ and $\hat s(l_i,d)$ also ensures the identifiability of the
GEV parameters $\mu^\gen(l_i)$ and $\sigma^\gen(l_i)$. Further, note that the 
choice of $\hat m(l_i,d)$ as maximal mean of all the ensemble members is in 
complete accordance to the choice of $v_{\max}^\pred$ in Equation
\eqref{eq:vmaxpred} below.
\medskip

\begin{figure}
  \begin{minipage}{0.016\textwidth}
  \vspace{-10.2cm} \textbf{a}
  \end{minipage}
  \begin{minipage}{0.4725\textwidth}
  \includegraphics[width=7.5cm]{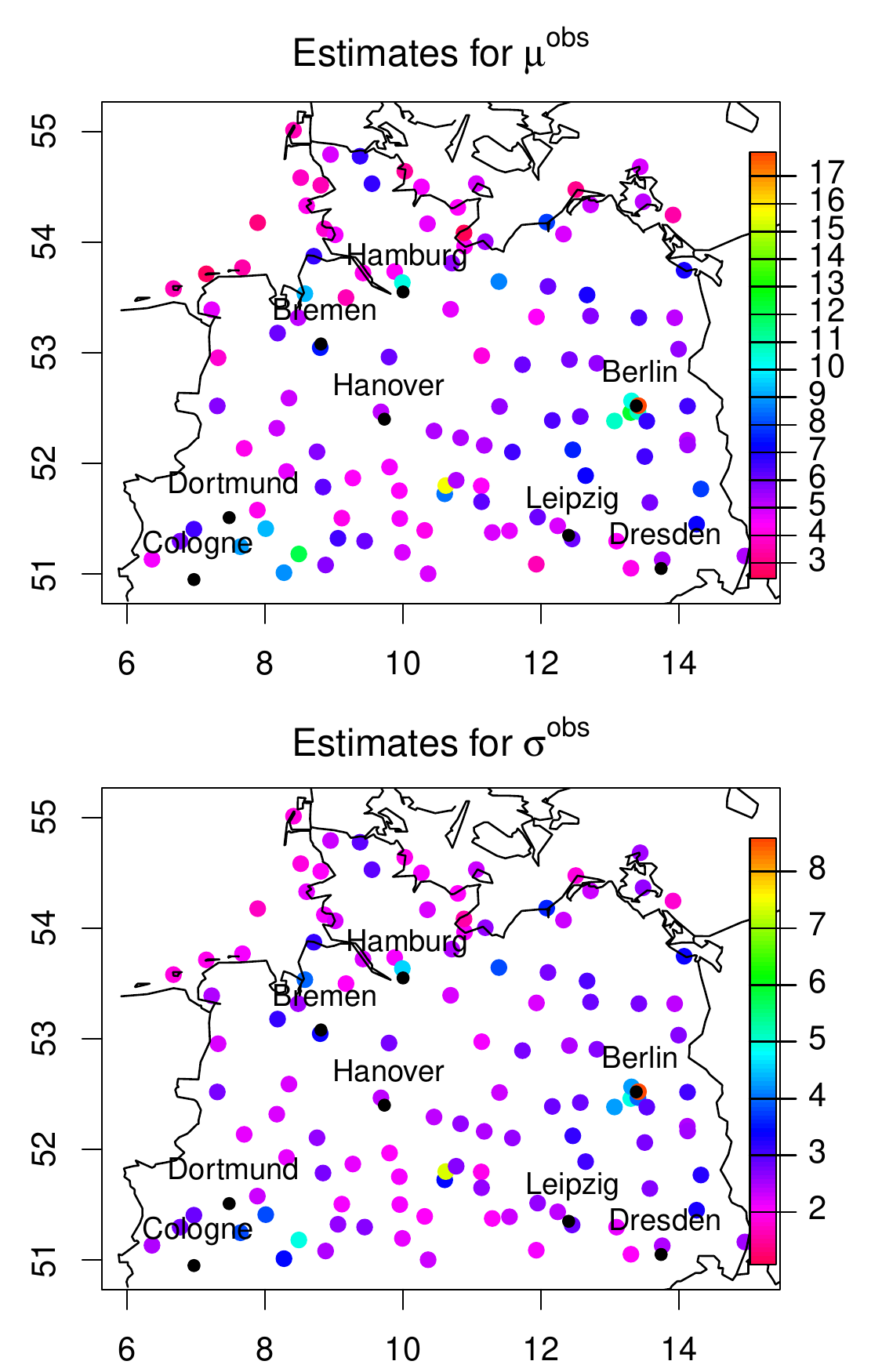}
  \end{minipage}
  \hfill
  \begin{minipage}{0.016\textwidth}
  \vspace{-10.2cm} \textbf{b}
  \end{minipage}
  \begin{minipage}{0.4725\textwidth}
  \includegraphics[width=7.5cm]{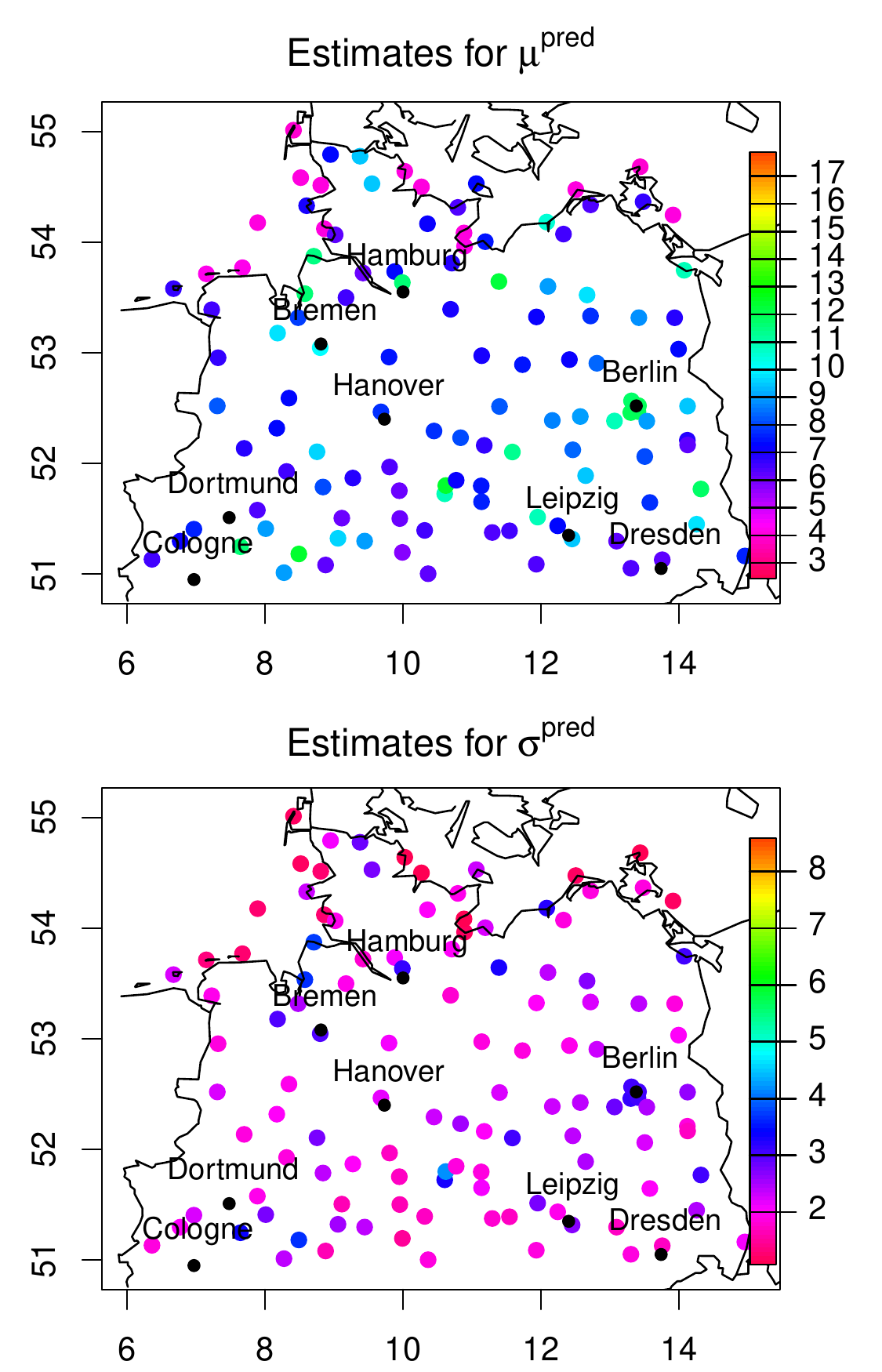}
  \end{minipage}
  \caption{\textbf{a} Estimates $\hat \mu^\obs(l_i)$ and
           $\hat \sigma^\obs(l_i)$ for the location and scale parameters
           corresponding to the observed maximal wind speed at the stations in
           the northern part of Germany.
           \textbf{b} Estimates $\hat \mu^\pred(l_i)$ and
           $\hat \sigma^\pred(l_i)$ for the location and scale parameters
           corresponding to the forecasted maximal wind speed at the stations in
           the northern part of Germany.}
  \label{fig:loc-scale-param}
\end{figure}

As described in Section \ref{sec:fit}, the GEV parameters for the standardized
observations can be estimated via maximum likelihood and the hypotheses that
these are spatially constant can be checked via Kolmogorov-Smirnov tests. For
$\xi^\obs$, we obtain a $p$-value of $0.194$. The analogous tests for 
$\mu^\obs$ and $\sigma^\obs$ both yield p-values smaller than 
$2.2 \cdot 10^{-16}$. Thus, the hypotheses that the residuals of the estimates
of $\mu^\obs$ and $\sigma^\obs$ follow a normal distribution both can be
rejected and, consequently, we drop the assumption that the GEV has the same
location and scale parameter at every station. In contrast, the shape parameter
of the GEV will be assumed to be spatially constant in northern Germany with 
the value $\xi^\obs = \hat \xi^\obs = 0.043$. Note, however, that the estimated
shape parameter differs significantly (to a $5\%$-level) from the mean value in 
case of 20 stations. For six of these stations, it even differs highly 
significantly (to a $1\%$-level), and four of them even to a $0.1\%$-level. The
parameter estimates $\hat \mu(l_i)$ and $\hat \sigma(l_i)$, $1 \leq i \leq 119$
for the location and scale parameters, respectively, obtained by maximum 
likelihood estimation with fixed shape parameter $\xi^\obs=\hat\xi^\obs$ are 
depicted in Figure \ref{fig:loc-scale-param}a. Note that the estimated vectors
of location and scale parameters show a strong empirical correlation of 0.97.
By \eqref{eq:obs-gumbel}, the data can be transformed to standard Gumbel 
margins. Kolmogorov-Smirnov tests performed separately for each station yield
$p$-values of at least 0.098 with a mean value of 0.718 which indicates that 
the GEV model fits quite well for all the stations.
\medskip

As a fit of the GEV distribution to the forecast is needed for both 
verification of the marginal model and the bivariate Brown-Resnick model, we 
repeat our analysis replacing the observed maximal wind speed
$v_{\max}^\obs(l_i,d)$ by $v_{\max}^\pred(l_i,d)$, i.e.\ a forecast for the 
maximal wind speed at station $l_i$ and day $d$. Here, we use the maximum over
the 20 corresponding COSMO-DE ensemble members
\begin{equation} \label{eq:vmaxpred}
 v_{\max}^\pred(l_i, d) = \max_{j=1, \ldots, 20} v_{\max}^{(j)}(l_i,d), \quad 1 \leq i \leq 119, \ 1 \leq d \leq 360,
\end{equation}
which ensures that the distribution of $v_{\max}^\pred$ is close to a GEV 
distribution.

As the Kolmogorov-Smirnov test of the normalized estimates for $\xi^\pred$
yields a $p$-value of $0.53$ and the estimates differ significantly from the
mean for seven stations (for three of them very significantly), we may assume
a shape parameter of $\xi^\pred = \hat \xi^\pred = 0.028$ at every station in
Northern Germany. However, the hypotheses that the estimates for the location 
and the scale parameter follow a normal distribution have been both rejected.
The maximum likelihood estimates $\hat \mu^\pred(l_i)$ and 
$\hat \sigma^\pred(l_i)$, $1 \leq i \leq 119$, with fixed shape parameter 
are shown in Figure \ref{fig:loc-scale-param}b. Here, the empirical correlation
of the vectors of estimated location and scale parameters is just as strong as 
in case of the observations. Kolmogorov-Smirnov tests of the transformed data
$x^\pred(l_i,d)$ for every station yield $p$-values of at least 0.142 with and
equal 0.748 in average which also indicates an appropriate fit.
\medskip

\begin{figure}
  \centering\includegraphics[width=12.6cm]{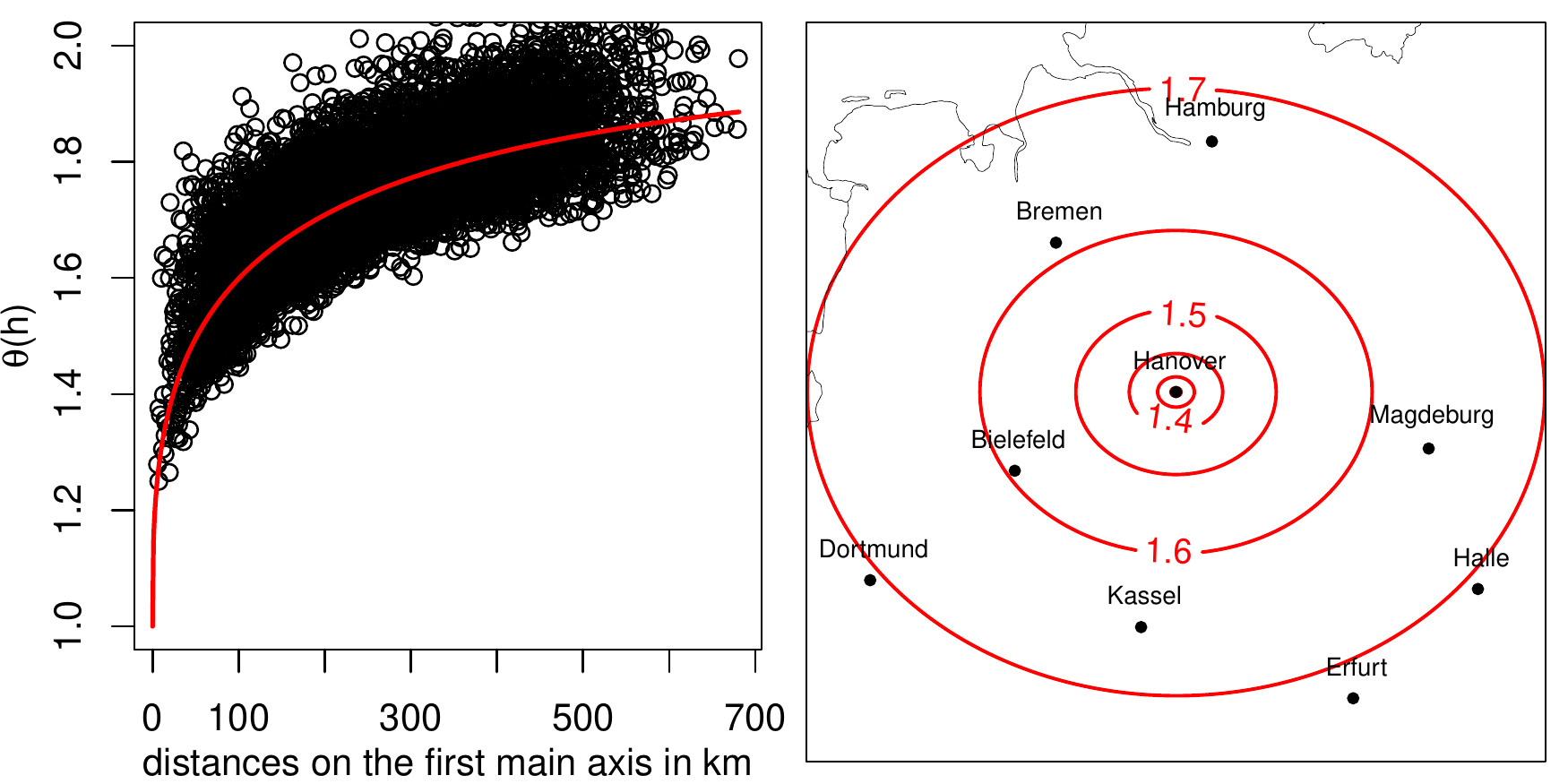}
  \caption{Left: The estimated extremal coefficients $\hat \theta^\obs$ (black
           circles) and the fitted extremal coefficient function
           $\tilde \theta^\obs$ (red line) of the normalized random field
           $X^\obs(\cdot,d)$ of observed wind gusts. Right: Contour level plot
           of the fitted extremal coefficient function
           $\tilde \theta^\obs(l_0, \cdot)$ where $l_0$ is located at Hanover.}
 \label{fig:obs-ecf}
\end{figure}

The spatial dependence is modeled by a univariate Brown-Resnick
process which is obtained by a weighted least squares fit of the extremal
coefficient function. Here, the weights depend on the variance of the
estimators $\hat \vartheta^\obs(l_i,l_j)$ (see Section \ref{sec:fit})
estimated by a jackknife procedure where the extremal coefficients are 
reestimated leaving out one month of data. The estimated extremal
coefficients $\hat\theta^\obs$ and the fitted extremal coefficient function
$$ \tilde\theta^\obs(s,t) = 2\Phi\left(\sqrt{\gamma_{\hat \vartheta^\obs}(s-t) / 2}\right), \quad s,t \in \RR^2.$$
are displayed in Figure \ref{fig:obs-ecf}. Here, the estimated coefficients
seem to be fitted quite well.

\medskip

For verification, we first calculate the mean CRPS for each of the two models
given by \eqref{eq:gen-model}, $\CRPS^{\obs}(l_i)$ and $\CRPS^{\pred}(l_i)$, 
for every station $l_i$, $1 \leq i \leq 119$. Then, the improvement or 
deterioration by using the GEV distributions of the observations instead of the
forecasts is expressed in terms of the skill score
\citep[e.g.,][]{gneiting-raftery-2007}
$$ S(l_i) = 1 - \frac{\CRPS^{\obs}(l_i)}{\CRPS^{\pred}(l_i)}$$
which has the value $1$ in case of an ``optimal'' model which equals 
$v_{\max}^\obs$ a.s.\ and the value 0 if both models yield the same result. 
Here, $S_{l_i} > 0$ for $115$ of $119$ stations. For the skill score 
corresponding to the mean $\CRPS$ averaged over all the stations, we obtain
$$S= 1 - \frac{\sum_{i=1}^{119} \CRPS^{\obs}_{l_i}}{\sum_{i=1}^{119} \CRPS^{\pred}_{l_i}} \approx 0.293.$$
Note that, for simplicity, the reference model \eqref{eq:gen-model} for the
predictions is based on the maximal ensemble members $v_{\max}^\pred(l_i,d)$
only and further information given by the maximal wind speed forecasted by the
other ensemble members are neglected. Thus, we further compare the CRPS of the
GEV model for the observations, $\CRPS^{\obs}(l_i)$, with the CRPS of the original
COSMO-DE ensemble, $\CRPS^{{\rm orig}}(l_i)$, taking the ensemble forecast as a 
probabilistic forecast with equal probability for each ensemble member.
Here, the skill $\tilde S(l_i) = 1 - \frac{\CRPS^{\obs}(l_i)}{\CRPS^{{\rm orig}}(l_i)}$
is positive for $37$ of $119$ only, with the skill of the averaged CRPS being
$1 - \frac{\sum_{i=1}^{119} \CRPS^{\obs}(l_i)}{\sum_{i=1}^{119} \CRPS^{{\rm orig}}(l_i)} \approx -0.032$.
As the skill score is slightly negative, the COSMO-DE ensemble forecast seems to 
contain more information than our marginal model. 
\medskip

For the verification of the spatial model, for all pairs of locations 
$(l_i,l_j)$, $1 \leq i < j \leq 119$, we estimate the energy scores 
$\widehat {ES}^{\rm BR}(l_i,l_j)$, based on $500$ samples of a Brown-Resnick
process, and compare them with the estimated scores 
$\widehat {ES}^{\rm ind}(l_i,l_j)$ for the independence model, based on 50 
samples of each GEV distribution. We obtain a positive skill score for $5819$
of $7002$ pairs of stations $(l_i,l_j)$  with a skill score of $0.025$ related
to the mean energy score. Although this improvement by the univariate Brown-Resnick 
model compared to the independence model in terms of predictive skill seems negligible,
realizations of gust fields look more realistic if spatial dependencies are respected.

Note that, for a fair comparison, we should avoid that training and validation
of the model are based on the same data. Hence, we perform cross validation
where the parameters are reestimated for every month, by leaving out the data
for this month and using only the data for the other months for training. The
GEV parameters estimated for different months in this way show very little variation
corroborating the assumption that they are constant in time. Further, the verification
results above are confirmed: We obtain skill scores of $0.285$ in the CRPS case
compared with the GEV model for the forecast, $-0.048$ compared to the COSMO-DE
ensemble and $0.035$ in case of the bivariate energy scores.

\subsection{Applying the Bivariate Model} \label{subsec:apply-biv}

A bivariate Brown-Resnick process is fitted to the transformed data according 
to Section \ref{sec:fit}. The cross-variogram parameter estimate 
$\hat\vartheta$ leads to the fitted extremal coefficient function
\begin{align*}
 \tilde \theta (l_i, l_j) ={} \left(\tilde \theta^{k_1,k_2}(l_i,l_j)\right)_{k_1, k_2 \in \genset}
={}& 2\left(\Phi\Big(\sqrt{\gamma_{k_1k_2}(\hat \vartheta;l_i-l_j)/2}\Big)\right)_{k_1, k_2 \in \genset}.
\end{align*}

\begin{figure}
 \centering\includegraphics[width=12.6cm]{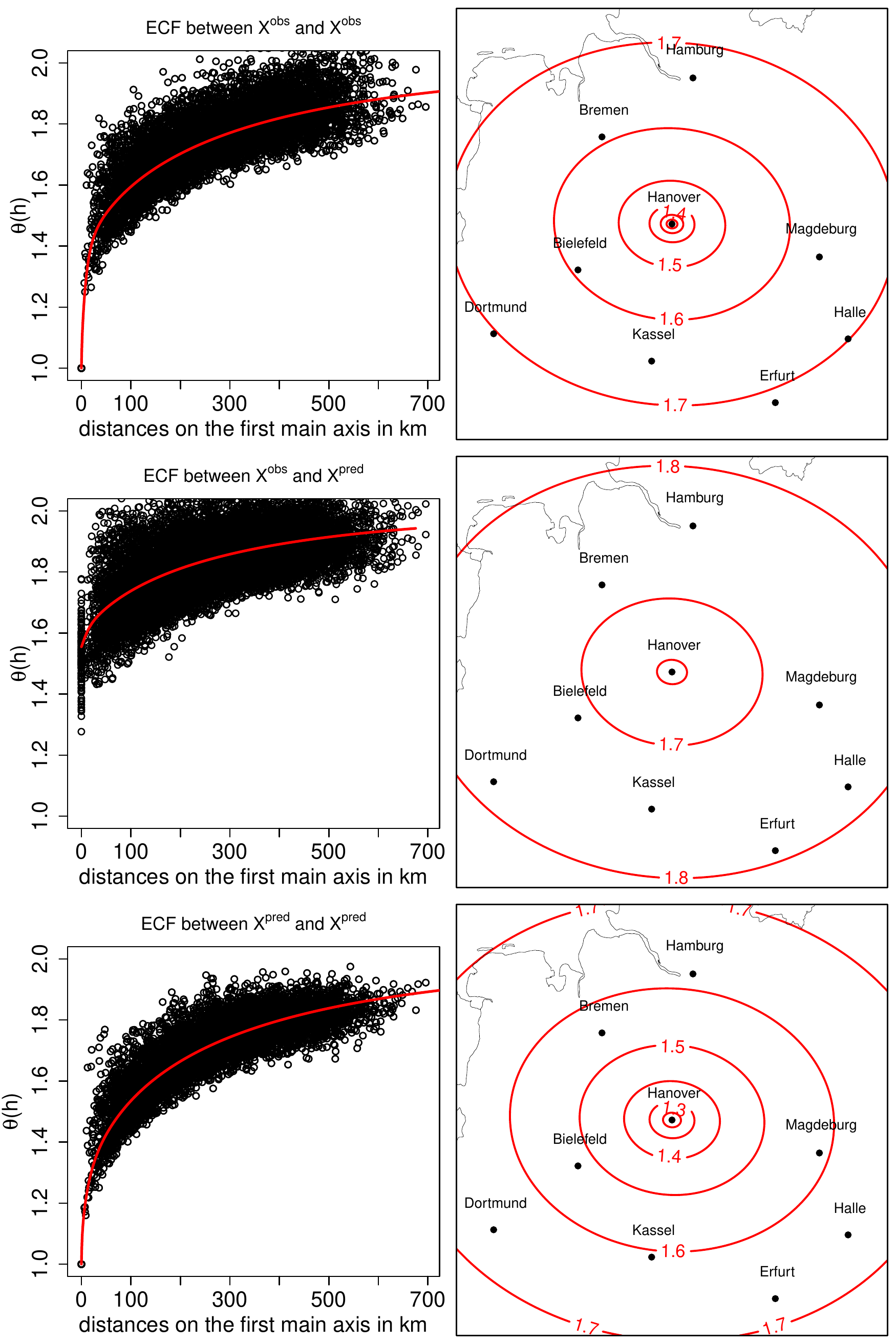}
  \caption{Left: The estimated extremal coefficients (black circles) and the
           fitted extremal coefficient function (red line) of the normalized
           bivariate random field $(X^\obs, X^\pred)$ of observed and forecasted
           wind gusts. Right: Contour level plots of the fitted extremal
           coefficient function $\tilde \theta(l_0, \cdot)$ where $l_0$ is
           located at Hanover.}
 \label{fig:all-ecf}
\end{figure}

\begin{figure}
  \centering\includegraphics[width=12.6cm]{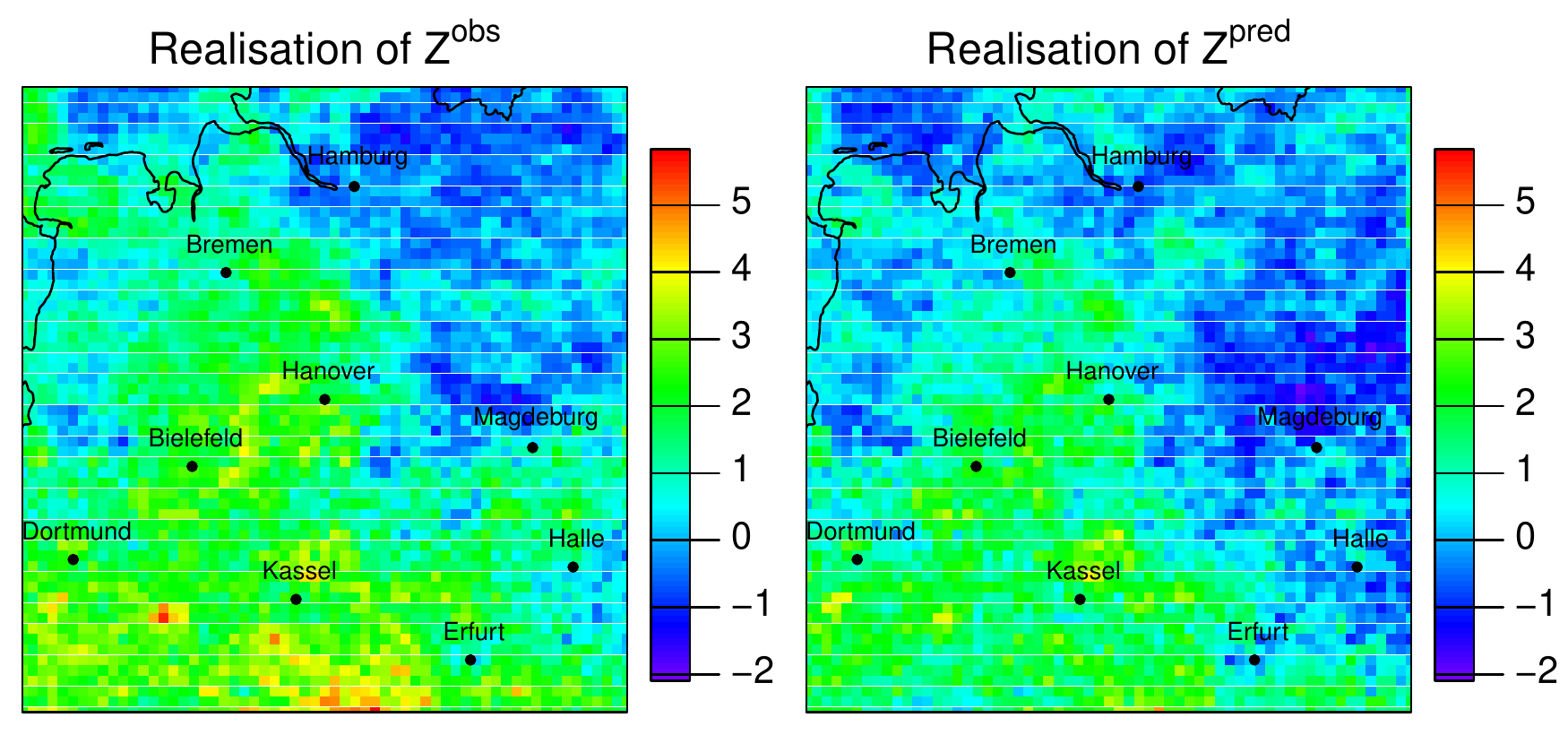}
  \caption{Simulated realization of a Brown-Resnick process associated to the
           variogram $\gamma(\vartheta; \cdot)$ with standard Gumbel margins.}
 \label{fig:biv-BR}
\end{figure}

Figure \ref{fig:all-ecf} presents the estimated extremal coefficients
$\hat\theta^{k_1,k_2}(l_i,l_j)$, and the fitted extremal coefficient functions
$\tilde\theta^{k_1,k_2}(\cdot,\cdot)$ for $k_1,k_2\in\genset$. As illustrated,
the fitted model seems to be appropriate with respect to the behavior of the
extremal coefficient function. 
Figure \ref{fig:biv-BR} depicts a simulated realization of the corresponding 
Brown-Resnick process associated to the variogram 
$\gamma(\hat \vartheta; \cdot)$ with standard Gumbel margins. The realization
indicates a remarkable amount of positive correlation between $x^\obs$ and
$x^\pred$ which emphasizes the gain of information by taking  $x^\pred$ into
account.

In order to verify the bivariate model, we apply the post-processing procedure
proposed in Subsection \ref{subsec:post}. However, due to the computational
complexity of the conditional simulation, we do not simulate the observations
at all stations simultaneously conditional on the forecast at all locations,
but perform post-processing with sample size $K=100$ at each location separately
conditioning on the forecast at the same location and two neighboring grid cells
only. We calculate the CRPS of the post-processed distribution, $\CRPS^{\rm biv}(l_i)$,
and compare it with $\CRPS^{(NWP)}(l_i)$, i.e.\ the CRPS belonging to the empirical
distribution of the original COSMO-DE ensemble, yielding a positive skill score
for $103$ of $119$ stations where the skill score related to the mean CRPS 
equals $0.164$ ($0.148$ cross-validated). Thus, we may conclude that the 
post-processing procedure based on the bivariate Brown-Resnick model is able to
improve the forecast given by COSMO-DE ensemble.

\section*{Appendix: Proof of Theorem \ref{thm:restriction}}
 For $i,j \in \{1,2\}$, and $h \in \RR^D$, we obtain
 \begin{align*}
  & \left( \sqrt{\gamma_{ii}(h)} - \sqrt{\gamma_{jj}(h)} \right)^2
   {}={} \gamma_{ii}(h) - 2 \sqrt{\gamma_{ii}(h) \gamma_{jj}(h)} + \gamma_{jj}(h) \displaybreak[0]\\
  \leq{}& \gamma_{ii}(h) - \Cov(W^{(i)}(h)-W^{(i)}(0), W^{(j)}(h)-W^{(j)}(0)) + \gamma_{jj}(h) \displaybreak[0]\\
  ={}& \frac 1 2 \Var\left(W^{(i)}(h) - W^{(i)}(0) - W^{(j)}(h) + W^{(j)}(0)\right) \displaybreak[0]\\
  ={}& \gamma_{ij}(0) - \Cov\left(W^{(i)}(h) - W^{(j)}(h), W^{(i)}(0)  - W^{(j)}(0)\right)
        + \gamma_{ij}(0) {}\leq{} 4 \gamma_{ij}(0),
 \end{align*}
 where we used the Cauchy-Schwarz inequality for both inequalities.
 Analogously, we get the assessment
 \begin{align*}
 & \left(\sqrt{\gamma_{ii}(h)} - \sqrt{\gamma_{ji}(h)}\right)^2
 {}={} \gamma_{ii}(h) - 2 \sqrt{\gamma_{ii}(h) \gamma_{ji}(h)} + \gamma_{ji}(h) \displaybreak[0]\\
  \leq{} & \gamma_{ii}(h) - \Cov(W^{(i)}(h)-W^{(i)}(0), W^{(j)}(h)-W^{(i)}(0)) + \gamma_{ji}(h) \displaybreak[0]\\
 ={}& \frac 1 2 \Var\left(W^{(i)}(h) - W^{(j)}(h)\right) {}={} \gamma_{ij}(0).
 \end{align*}
 Thus, the assertion of the theorem follows with $\gamma_0 = \gamma_{11}$.

\bibliographystyle{Chicago}
\bibliography{ref}
\end{document}